\documentclass[aps,twocolumn,prb,superscriptaddress,showpacs,preprintnumbers,amsmath,amssymb]{revtex4}

\usepackage{graphicx}
\usepackage{dcolumn}
\usepackage{bm}

\usepackage{color}

\newcommand{\In}{InCu$_{2/3}$V$_{1/3}$O$_{3}$}

\begin{document}

\preprint{\it to appear as Rapid Com.  in PRB (2010)}

\title{Finite size effects and magnetic order in the spin-1/2 honeycomb lattice compound \In}

\author{M. Yehia}
\affiliation{Leibniz Institute for Solid State and Materials Research IFW Dresden, D-01171 Dresden, PO BOX 270116, Germany}
\author{E. Vavilova}
\affiliation{Leibniz Institute for Solid State and Materials Research IFW Dresden, D-01171 Dresden, PO BOX 270116, Germany} \affiliation{Zavoisky
Physical Technical Institute of the Russian Academy of Sciences, 420029, Kazan, Russia}
\author{A. M\"{o}ller}
\affiliation{University of Houston, Department of Chemistry and Texas Center for Superconductivity, Houston, TX 77204, USA}\affiliation{Institut f{\"u}r
Anorganische Chemie, Universit{\"a}t zu K\"{o}ln, 50939 K\"{o}ln, Germany }
\author{T. Taetz}\affiliation{Institut f{\"u}r Anorganische
Chemie, Universit{\"a}t zu K\"{o}ln, 50939 K\"{o}ln, Germany }
\author{U. L\"{o}w}\affiliation{Technische Universit{\"a}t Dortmund, Theoretische Physik I, 44221 Dortmund, Germany}
\author{R. Klingeler}\affiliation{Leibniz Institute for Solid State and
Materials Research IFW Dresden, D-01171 Dresden, PO BOX 270116, Germany}
\author{V.Kataev}
\email{v.kataev@ifw-dresden.de}
\affiliation{Leibniz Institute for Solid State and Materials Research IFW Dresden, D-01171 Dresden, PO BOX 270116,
Germany}
\author{B. B\"{u}chner}\affiliation{Leibniz Institute for Solid State and
Materials Research IFW Dresden, D-01171 Dresden, PO BOX 270116, Germany}

\date{22 February 2010}

\begin{abstract}
High field electron spin resonance, nuclear magnetic resonance and magnetization studies addressing the ground state of the quasi two-dimensional
spin-1/2 honeycomb lattice compound InCu$_{2/3}$V$_{1/3}$O$_{3}$ are reported. Uncorrelated finite size structural domains occurring in the honeycomb
planes are expected to inhibit long range magnetic order. Surprisingly, ESR data reveal the development of two collinear antiferromagnetic (AFM)
sublattices below $\sim 20$\,K whereas NMR results show the presence of the staggered internal field. Magnetization data evidence a spin
reorientation transition at $\sim 5.7$\,T. Quantum Monte-Carlo calculations show that switching on the coupling between the honeycomb spin planes in
a finite size cluster yields a N\'{e}el-like AFM spin structure with a substantial staggered magnetization at finite temperatures. This may explain
the occurrence of a robust AFM state in \In\ despite an unfavorable effect of structural disorder.
\end{abstract}






\maketitle

In planar honeycomb lattice systems a combination of nontrivial topology, strong electronic, spin and orbital correlations and degeneracies yields a
rich variety of ground states, novel excitations and exotic behaviors  that currently attract much attention. The recently discovered exciting
phenomena range, e.g., from the quantum Hall effect in graphene \cite{Novoselov05,Du09}, superconductivity in MgB$_2$ \cite{Nagamatsu01} and
intercalated graphite \cite{Weller05}, to topologically driven quantum phase transitions in anyonic quantum liquids \cite{Gils09}.

Regarding the spin degrees of freedom, an important feature of low dimensional spin systems is the presence of quantum fluctuations that inhibit long
range order of the quantum spin-1/2 lattice. Such an effect essentially depends on the spin coordination number $z$. The one dimensional (1D)
Heisenberg antiferromagnetic (AFM) $S = 1/2$ chain with $z = 2$ does not show any magnetic order even at zero temperature \cite{Anderson52}, whereas
long range order is possible at $T = 0$ in the 2D case \cite{MWH} as, e.g., in the prominent $S = 1/2$ Heisenberg square lattice model with $z = 4$.
The honeycomb lattice has the minimum possible coordination number of any regular 2D lattice $z = 3$. Thus quantum fluctuations there are weaker than
in the 1D case, but stronger than in the 2D square lattice. Hence, the AFM order for the honeycomb lattice is fragile \cite{Bishop1998,Fouet2001}.

Experimentally, low-D spin systems described by the Heisenberg Hamiltonian $\mathcal{H} = 2J_{\rm afm}\sum S_iS_j$ are often realized in structurally
three dimensional organic- or transition metal oxide (TMO) compounds where strong AFM exchange interaction $J_{\rm afm}$ between unpaired localized
spins occurs along only one or two spacial directions. Owing to residual small 3D exchange couplings such materials usually exhibit a long range
N\'{e}el order at a finite temperature $T_{\rm N}$, though, unlike in the 3D magnets, the ordering occurs at much smaller temperatures $T_{\rm N}\ll
J_{\rm afm}/k_{\rm B}$.

Recently, a complex TMO compound, \In, was suggested as a possible candidate for the realization of the $S = 1/2$ honeycomb lattice
\cite{Kataev2005}. In its layered hexagonal structure the Cu$^{2+}$ ($3d^9, S = 1/2$) ions are proposed to be arranged in a 2D network of hexagons
with the nonmagnetic V$^{5+}$ ($3d^0$) ions in the center of each hexagon. The honeycomb layers are separated by sheets of [InO$_6$] polyhedra along
the hexagonal $c$ axis. The analysis of the static susceptibility  $\chi(T)$ has yielded the exchange parameter $J_{\rm afm}\simeq 140$\,K of the
honeycomb spin lattice \cite{Kataev2005}. A later structural neutron diffraction study revealed that a structural $_\infty ^2 $\{V$_1$Cu$_{6/3}\}$
order in the hexagonal planes has a finite correlation length $\xi_{\rm st} \sim 300 \AA$ and that these structural domains are randomly arranged
along the $c$ axis \cite{Moller2008}. As has been argued in Ref.~\onlinecite{Moller2008} this structural disorder rendering finite values of the
spin-spin correlation length $\xi_{\rm s} \leq \xi_{\rm st}$ concomitant with the substantial spacing between the honeycomb planes and with the low
spin coordination number in the planes could preclude a 3D N\'{e}el order in \In. Thus the kink-like anomalies in $\chi(T)$-dependence at 30\,K and
38\,K previously identified with the transition to the long range ordered state \cite{Kataev2005} have been tentatively assigned to glass-like order
of unsaturated spins in domain boundaries \cite{Moller2008}.

To obtain insights into the nature of the ground state of \In\ we have investigated its magnetic properties using two local spin probe techniques,
high field electron spin resonance (ESR) and nuclear magnetic resonance (NMR). These experiments were complemented by measurements of the field
dependence of the static magnetization $M(B)$ and by quantum Monte-Carlo calculations. From ESR and NMR data we have obtained evidence for the
development of the AFM sublattices in the spin system below $\sim 20$\,K which according to the $M(B)$ data experience a reorientation transition at
a field of $\sim 5.7$\,T. Our Monte-Carlo calculations reveal that switching on the interlayer coupling in a finite-size honeycomb spin domain yields
the formation of the staggered sublattices with substantial sublattice magnetization at a finite temperature. We conclude that structural disorder
does not inhibit the N\'{e}el ordered state in \In\ at least on a spacial scale of structural domains which, owing to their large size, has
properties very similar to the long range AFM order in the infinite systems.

Clean stoichiometric powder of \In\ was synthesized and thoroughly characterized as described in Ref. \onlinecite{Moller2008}. From it we prepared an
'oriented' sample by mixing the powder with epoxy and letting it harden in a magnetic field of 1\,T. Owing to the specific anisotropy of the
Cu$^{2+}$ $g$ factor \cite{Kataev2005} the $c$ axis of the powder particles was oriented perpendicular to the orientation axis (hereafter $o$ axis)
defined by the direction of the applied field, as confirmed by x-ray diffraction experiments. ESR was measured with a home made spectrometer on the
basis of the Millimeterwave Vector Network Analyzer from AB Millimetre, Paris, combined with a 17\,T superconducting magnetocryostat from Oxford
Instruments Ltd.~\cite{Golze2006}. NMR data were collected on a Tecmag pulse solid-state NMR spectrometer with a 9.2\,T superconducting magnet from
Magnex Scientific. The NMR spectra were obtained by measuring the intensity of the Hahn echo versus magnetic field. The $T_1$ relaxation time was
measured with the method of stimulated echo. Static magnetization was measured with a superconducting quantum interference device  magnetometer from
Quantum Design. In all experiments the magnetic field was applied parallel to the $o$ axis, i.e., perpendicular to the hexagonal $c$ axis.

In the previous ESR experiment performed at a standard X-band 'low' frequency $\nu = 9.47$\,GHz a strong paramagnetic Cu$^{2+}$ resonance signal
clearly visible at high temperatures experienced at $T< 80$\,K strong broadening and shift and could not be detected below 50\,K \cite{Kataev2005}.
Since such a behavior could be indicative of an opening of the energy gap for resonance excitations, e.g., due to the establishment of the AFM order,
we have measured in the present work the ESR spectra at low $T$ in a broad frequency domain extending into the sub-THz range.

\begin{figure}
\includegraphics[width=0.70\columnwidth,clip]{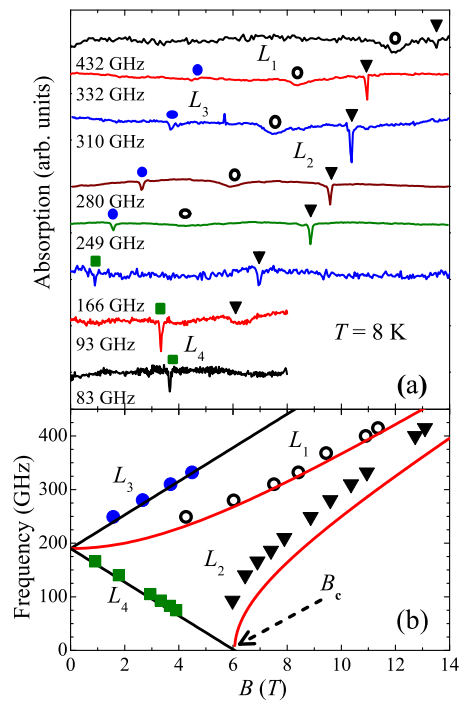}
\caption{(Color online) Frequency $\nu$ dependence of the ESR signals at $T = 8$\,K: (a) Selected spectra at different $\nu$. Open circles, solid
circles, triangles and squares indicate the resonance field $B_{\rm res}$ of the lines $L_1$ - $L_4$; (b) The $\nu$ vs. $B$ diagram of the resonance
modes. The symbols correspond to respective $L_{\rm i}$ in panel (a), solid lines are model curves (see the text).} \label{ESR}
\end{figure}

ESR spectra at various frequencies $\nu$ collected at $T = 8$\,K for a magnetic field applied parallel to the orientation axis  ($B\| o$ axis) are
shown in Fig.~\ref{ESR}(a). Several resonance modes are visible. At high frequencies three  modes can be detected, $L_{1}$, $L_{2}$ and $L_{3}$. They
shift towards higher fields by increasing the $\nu$. At lower frequencies, the $L_{4}$ mode can be observed which, in contrast, moves towards lower
fields by increasing the $\nu$. The evolution of $L_1$ - $L_4$ is summarized in the frequency-field ($\nu$ vs. $B$) diagram in Fig.~\ref{ESR}(b).
Notably, none of the $\nu(B)_{L{\rm i}}$ branches correspond to a paramagnetic resonance condition $h \nu =g \mu_{B}B_{\rm res}$. Here $h$ is the
Planck constant and $\mu_B$ is the Bohr magneton. Modes $L_1$, $L_3$ and $L_4$ are obviously gapped. Extrapolation of $L_3$ and $L_4$ branches to
zero field yields an estimate of the gap value $\Delta \simeq 190$\,GHz. Mode $L_4$ softens with increasing the field strength revealing a critical
field $B_{\rm c} \sim 5.5 - 6$\,T where $\nu_{L{\rm 4}}\rightarrow 0$. Above $B_{\rm c}$ mode  $L_2$ emerges. In fact, taking the above estimate of
$\Delta$ and the value of the $g$ factor for $B\| o$ axis $g = 2.23$ obtained from the X-band measurements at high $T$ (not shown) it is possible to
reasonably model the $L_1 - L_4$ branches with the AFM resonance (AFMR) relations for a uniaxial two-sublattice antiferromagnet \cite{Turov}. $L_{3}$
and $L_{4}$ can be represented as $h \nu = \Delta \pm g \mu_{B} B_{\rm res}$, where the plus and minus signs correspond to $L_{3}$ and $L_{4}$,
respectively. $L_2$ that appears above $B_{\rm c}$ can be described as $h \nu / g \mu_{B}= \sqrt{B_{\rm res}^{2} - \Delta^{2}}$. Finally, $L_1$
follows the relation $h \nu / g \mu_{B}= \sqrt{\Delta^{2}+ B_{\rm res}^{2}}$. Apart from some systematic discrepancy for the $L_2$ branch the model
agrees well with the experiment. The behavior of $L_2$, $L_3$ and $L_4$ is typical for the situation when a magnetic field is applied parallel to
some 'easy' axis of an antiferromagnet whereas $L_1$ corresponds to a 'hard' direction \cite{Turov}. The field $B_{\rm c}$ at which for $L_{\rm 2}$
and $L_{\rm 4}$ the frequency $\nu \rightarrow 0$ is the spin-flop field $B_{\rm c}$ related to $\Delta$ as $B_{\rm c} = (h/g\mu_B)\Delta$
\cite{Turov}. Taking $\Delta \simeq 190$\,GHz one obtains $B_{\rm c} \sim 6$\,T. Since $B$ was applied in the honeycomb plane of \In, there must be
therefore some 'easy' direction in the plane that defines the AFM vector of two sublattices. Obviously the direction of this in-plane 'easy' axis of
each powder particle is random in the sample, because specific in-plane crystallographic directions in our sample are not defined. Therefore both
'easy' and 'hard' AFMR modes are present. We note that these modes can be seen at temperatures up to $\sim 20$\,K.

\begin{figure}
\includegraphics[width=0.8\columnwidth,clip]{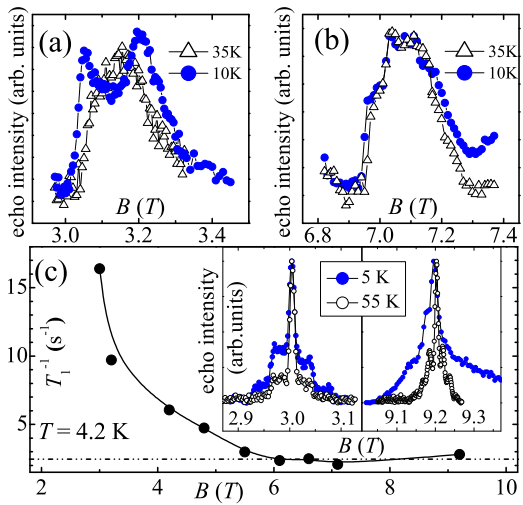}
\caption{(Color online) $^{114}$In  NMR line at 35\,K and 10\,K at fields smaller (a) and larger (b) than $B_{\rm c} \sim 6$\,T; (c) $T_1^{-1}$
relaxation rate of $^{51}$V nuclei as a function of $B$ at $T = 4.2$\,K. Lines are guides for the eye. Inset: $^{51}$V NMR spectra at fields smaller
(left) and larger (right) than $B_{\rm c}$ at 55\,K and 5\,K.} \label{NMR}
\end{figure}

\begin{figure}
\includegraphics[width=0.75\columnwidth,clip]{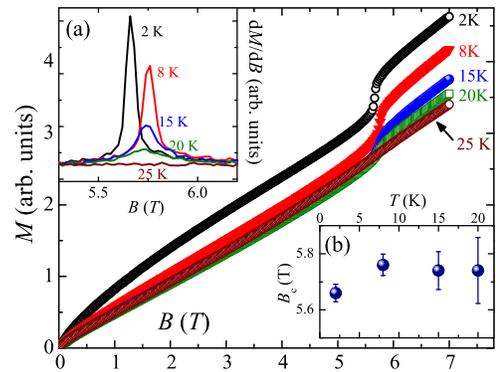}
\caption{(Color online)Field dependence of the magnetization $M(B)$ at different temperatures for $B \|$ o-axis after a subtraction of a small
$T$-independent spurious ferromagnetic-like signal persisting up to room temperature. Inset: (a) - field derivative of $M(B)$; (b) $T$ dependence of
the reorientation field. Error bars correspond to the width of the transition.}\label{magn}
\end{figure}

Further evidence for the occurrence of the AFM sublattices and related staggered magnetization has been obtained from the NMR measurements of
$^{51}$V $(I = 7/2)$ and  $^{114}$In $(I = 9/2)$ nuclei in \In. At high temperatures the spectra of both nuclei consist of a main line and a typical
satellite structure arising due to the quadrupole interaction. In is positioned outside the hexagonal (Cu,V)-plane asymmetrically with respect to the
Cu honeycomb ring. Hence one could expect at low $T$ an uncompensated staggered field at this position. Indeed, below 20\,K, where AFMR modes are
observed in the ESR experiment, the $^{114}$In NMR signal experiences a remarkable change if measured in a magnetic field $B_{\rm NMR}$ smaller than
$B_{\rm c}$ [cf. Figs.~\ref{NMR}(a)and (b))]. At $T = 35$\,K the line is unsplit regardless the field of the measurement. However, at $T = 10$\,K the
signal splits in two components for $B_{\rm NMR} < B_{\rm c}$ whereas it remains unsplit for $B_{\rm NMR} > B_{\rm c}$. This observation strongly
supports the scenario of compensated collinear AFM sublattices at $B < B_{\rm c}$ which turn into the spin flop state at $B > B_{\rm c}$. The
$^{51}$V NMR data agree with this scenario. The V site is symmetric with respect to the Cu sites. Therefore the staggered fields from the honeycomb
spin lattice should be compensated at this position. In fact, the $^{51}$V NMR spectrum does not change qualitatively with decreasing the $T$ at
$B_{\rm NMR} < B_{\rm c}$ and shows the presence of a broad background at a high field [Fig.~\ref{NMR}(c), inset]. The absence of the broadening of
the main line suggests that this background is due to the influence of structural domain walls where uncompensated spins create a distribution of the
local magnetic fields. Remarkably, though the $^{51}$V NMR signal remains unsplit at low $T$, the $^{51}$V longitudinal nuclear relaxation rate
 $T_1^{-1}$ shows a substantial field dependence: It decreases by a factor of 7
by approaching the $B_{\rm c}$ and shows no field dependence at $B_{\rm NMR} > B_{\rm c}$ [Fig.~\ref{NMR}(c)].
An initial rapid drop of $T_1^{-1}$ at $B\sim 3$\,T could be due to a field-suppression of magnetic fluctuations caused by impurity free-like spins.
However, according to the magnetization measurements (see Fig.~\ref{magn} and its discussion below) these spins are fully polarized at stronger
fields. Thus a further substantial reduction of $T_1^{-1}$ is likely to be related with the approach to the critical field $B_{\rm c}$.
The sensitivity of  $T_1^{-1}$ to the spin structure may be related to magnetic anisotropy that defines a specific 'easy' orientation in the
honeycomb spin plane and possibly produces a small uncompensated fluctuating field at the V nuclei in the collinear phase that become ineffective for
the $T_1$ relaxation after the spin reorientation has occurred.

Motivated by the findings of the local ESR and NMR techniques we have performed the magnetization measurements on an oriented powder sample of \In\
for a magnetic field applied parallel to the $o$ axis. The temperature dependence of the magnetic susceptibility $\chi = M/B$ at $B=1$\,T
demonstrates the same behavior as observed before in Ref.~\onlinecite{Moller2008} (not shown). The field dependence of the magnetization $M(B)$ at
different $T$ is presented in Fig.~\ref{magn}. The curvature of the $M(B)$ at low fields is likely due to a small amount of  free-like spins, which
also yield a Curie-like tail in the $T$ dependence of $\chi$ (Ref.~\onlinecite{Moller2008}). The $M(B)$ dependence at the lowest temperature $T =
2$\,K exhibits a pronounced step-like increase of $M$ at $B \simeq 5.7$\,T. It can be straightforwardly interpreted as a critical field $B_{\rm c}$
for the spin reorientation, in a nice agreement with ESR and NMR data. By increasing the $T$ the step continuously broadens, slightly shifts towards
higher fields and finally smoothly vanishes above $\sim 20$\,K [Fig.~\ref{magn}, insets (a,b)].
\begin{figure}
\includegraphics[width=0.8\columnwidth,clip]{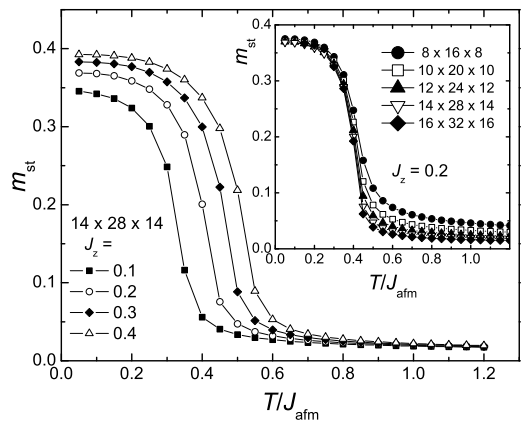}
\caption{$T$ dependence of the staggered magnetization $m_{\rm st}$ of the honeycomb lattice with interplane couplings $J_z = 0.1, 0.2, 0.3$ and
$0.4J_{\rm afm}$ for the system size $14 \times 28 \times 14$. Inset: $m_{\rm st}$ vs. $T/J_{\rm afm}$ for $J_z = 0.2 J_{\rm afm}$ for different
system sizes.} \label{QMC}
\end{figure}

Thus the results of three different techniques strongly suggest the formation of collinear AFM N\'{e}el sublattices in the $S=1/2$ honeycomb plane in
\In\ at a finite temperature.   As discussed in Ref.~\onlinecite{Moller2008} the absence of the magnetic anomaly in the specific heat $C_{\rm p}$
does not exclude the occurrence of the magnetic phase transition in a low-dimensional system such as \In\  due to residual interactions in the third
dimension. If the ordering temperature $T_{\rm N}$ is substantially smaller than $J_{\rm afm}/k_{\rm B}$, most of the magnetic entropy $S_{\rm mag}$
is already gone at $T \sim J_{\rm afm}/k_{\rm B}$ due to the 2D in-plane AFM correlations. Thus an additional change of $S_{\rm mag}$ at $T_{\rm N}$
might be small and not visible in the $C_{\rm p}$ owing to a substantial phononic background. The question arises if this 3D situation is realized in
\In\ despite the occurrence of disordered finite size structural domains. To address this issue we have performed a computational study of a finite
size spin-1/2 clusters on the honeycomb lattice by means of a continuous Euclidean time Quantum-Monte-Carlo algorithm that includes the coupling in
the third dimension \cite{Loew09}. We find that the N\'{e}el type structure and the staggered magnetization $m_{\rm st}$, otherwise present in the 2D
case only at $T = 0$, develops in a rather narrow interval at a finite temperature that depends on the strength of the interlayer coupling $J_z$. As
shown in the main panel of Fig.~\ref{QMC} a temperature interval where $m_{\rm st}(T)$ exhibits a step-like increase is mainly determined by the
value of $J_z$ whereas the sharpness of this step increases with the increase of the cluster size (inset of Fig.~\ref{QMC}). Note that in \In\ the
size of in-plane structural domains amounts to $\sim 300 \AA$ (Ref. \onlinecite{Moller2008}) which corresponds to $\sim 90 \times 90$ spin sites,
i.e. a much larger number than in our model calculation. Thus one should expect quite a sharp set up of the two-sublattice collinear AFM N\'{e}el
spin structure and the occurrence of the staggered magnetization in \In\ that experimentally would be difficult to discriminate from a true magnetic
phase transition expected in the thermodynamic limit. Finally, regarding modelling of the finite size systems, we mention here that a calculation in
the framework of the (anisotropic) Heisenberg model of the ESR response of the AFM cluster comprising 8 spins only already yields AFMR modes of a
substantial intensity that for the 'easy' axis direction have qualitatively similar properties as those shown in Fig.~\ref{ESR} \cite{Ogasahara00}.

In summary, by studying ESR, NMR and static magnetization of the oriented powder sample of the honeycomb spin-1/2 lattice compound \In\ we have found
strong experimental evidence for the formation of the N\'{e}el type collinear AFM spin structure at temperatures below $\sim 20$\,K and a respective
development of the staggered magnetization. A reorientation of spin sublattices in a field of $\sim 5.7$\,T has been clearly identified in the
magnetization and magnetic resonance data. The AFM ordering of spins in InCu$_{2/3}$V$_{1/3}$O$_{3}$ takes place despite the low spin coordination
number, the pronounced structural two dimensionality as well as the presence of the finite size incoherent structural domains. Our experimental
results are corroborated by the Quantum Monte-Carlo study of the finite size honeycomb spin layers coupled in the third dimension. It reveals a sharp
development of the staggered magnetization at a finite temperature that in the thermodynamic limit would correspond to a phase transition to the AFM
long-range ordered state.

The work was supported in part by the German-Russian cooperation project
of the DFG (grant No. 436 RUS 113/936/0-1) and by the Russian Foundation
for Basic Research (grant No. 08-02-91952-NNIO-a) . M.Y. acknowledges the
DAAD for a scholarship, and T.T. acknowledges the DFG(SFB608) for support.


\end{document}